\documentstyle[12pt]{article}
\def\frac#1#2{{\displaystyle#1\over\displaystyle#2}}

\begin{document}
\begin{center}
{\bf GAMMA RADIATION BY NEUTRALINO STARS}
\footnote{Published in Phys.Lett.A {\bf 225}, 217 (1997)}

\vspace{5mm}

A.V.Gurevich and K.P.Zybin\\
\vspace{5mm}
{\em  P.N.Lebedev Institute of Physics, Russian Academy of Sciences,
Moscow, Russia}\\
\end{center}
\begin{abstract}
A new model of gamma radiation (GR)  generated by neutralino
annihilation in neutralino stars (NeS) is proposed. Diffuse Galactic
and extragalactic  GR is calculated for this model and is shown to be
in a reasonable agreement with observations.

The point source component (P) is picked out  among nonidentified
discrete gamma sources from EGRET catalog. This component is shown
to be distributed isotropically and homogeneously, as it should be
for NeS. Gamma radiation from significant part of
P-sources is supposed to be the direct radiation
from individual NeS. This supposition is shown to be in  agreement
with observations. The comparison of the theory with observed GR gives
a possibility to establish some important features of CDM particle.
\end{abstract}
\section{Introduction}
Microlensing observations established, that the large part of
dark matter in Galaxy halo consist of unseen objects with masses
$(0.05-0.8)M_{\odot}$ [1,2]. These objects are naturally considered
to be White Dwarfs, Brown Dwarfs, Jupiters or some other stars
or planets. To the contrary in [3,4] was supposed the existence of
quite a new type of massive structures -- gravitationally
compressed objects from nonbarionic cold dark matter (CDM).
These objects are the results of a small scale hierarchical structuring
developing in CDM [5,6]. They are noncompact, having some spectrum of
characteristic dimensions and masses.

After recombination a
barionic component falls down into the potential well formed by
nonbarionic matter. In this way the compact barionic core
is created. The nonbarionic component forms a
spread spherical halo, where the dominant mass of the object
is concentrated. The main hypothesis of [3]-[6] is that these
noncompact nonbarionic objects determine significant part of
microlensing events observed in [2].

The particles of which CDM is composed are not known now, though
there exist some hypothetical candidates: neutralino, heavy
neutrino, axions, strings. Neutralino and heavy neutrino like
the Majorana particles can annihilate in mutual collisions.
Noncompact objects which consist of such particles are
partially disappearing
 because of annihilation processes. These specific objects
we call "neutralino stars" (NeS) [3] and for nonbarionic particles
from which they consist of we will use
as a general name "neutralino" [4].

Annihilation of neutralino leads to effective radiation of gamma
photons,
heating of barionic core etc. It should be emphasized, that the
energy released in all this processes was input into CDM during
the period of its freezing out. The possibility to detect CDM
particles through their annihilation was discussed in a number of
papers (see [7,8]). In these papers always was supposed a smooth
distribution of CDM in the Galaxy.
The existence of NeS lead to a new situation
, connected with compression of CDM in NeS, which
results in strong amplification
of neutralino annihilation processes. It opens from one side  the
 possibility of existence of quite observable fluxes of gamma radiation
by NeS and from the other side
gives significant constrains  on the possible type of neutralino
particles.

Calculations of the gamma radiation produced by NeS and comparison
of theoretical results with observational data is the goal of the
present paper. In section 2 the structure of small-scale
noncompact objects (NeS) is considered. In section 3 the
flux of  gamma radiation produced by NeS is discussed.
Diffusive Galactic and extragalactic gamma fluxes are calculated.
Discreet gamma sources are discussed as well. In section 4 predictions
of the theory are compared with observations.
The results of observations of diffusive fluxes and recently discovered
new discreet gamma sources  are shown to be in a reasonable agreement
with the theory. It allows  us to propose  special
experiments, using measurements of gamma radiation  from discreet
sources which could result in discovery of neutralino stars and may
indicate the type of CDM particle.

\section{Structure of small scale dark matter objects}

Gamma emission  depends essentially on the distribution of
neutralino density $n_x(r)$ in NeS. The last one is determined by nonlinear
process of Jeans gravitational collapse in nondissipative cold dark
matter. This process was the subject of multiple studies [9]. The
dynamical solution, using analytical approach [10], showed, that the
growth of Jeans instability at large scales results in the creation of a
stationary, spherically symmetrical objects. For these objects the
theory established a fundamental scaling law for the
density distribution [11]:
\begin{equation}
\rho\propto r^{-\alpha},\qquad \qquad \alpha\approx 1.8
\end{equation}
This law fits well the galaxies rotational curves [12]. Basing on the same
dynamical solution (1) a well-known scaling law was obtained for the
distance dependence of correlation functions for galaxies and other
large scale objects [13].

Scaling law (1) is the result of two main factors: long-time linear
growth of initial density fluctuations and fast nonlinear gravitational
collapse of nondissipative dark matter. The initial density fluctuations
for a large scale structures are always small
$\delta_i=\delta\rho/\rho <<1$ and because of this the unstable mode
is growing up for the long enough time, before it reaches the nonlinear
stage. The stable modes during the same time are effectively damping
and because of this up to the moment , when nonlinear process begins
they remain very small. But during nonlinear collapse stage one of
the dumping modes begin to grow and this growing goes very fast,
especially in the vicinity of the centrum ($r\to 0$) of the density
distribution (1) [10]. Under the influence of this mode the singular law
(1) cuts of on the scales:
\begin{equation}
r_c\approx \delta_i^3 R_x
\end{equation}
where $R_x$ is the characteristic size of the object. As the initial
value $\delta_i$ for the large scale structures is always small enough
$\delta_i\leq 10^{-3}$, the scaling law (1) can be fulfilled up to very
small values $r_c\leq 10^{-9} R_x$. In reality, the distribution (1)
changes  much earlier, being affected by the dissipative
barionic matter.

For the small scale structures situation is essentially different.
First  the noncompact objects of the type observed by microlensing
$M_x\sim(0.01-1)M_{\odot}, R_x\sim10^{14}-10^{15}$cm could develop
only if the initial density fluctuations are not small ($\delta_i
\sim0.3-0.5$) [4,5,14]. From (2) it follows that the nonbarionic core
radius $r_c$ in this case is:
\begin{equation}
r_c\sim 0.1 R_x.
\end{equation}
Second, the considered objects are generated as a result of a long
developed hierarchical structure [5,6]. The specific feature of this
structure is that the most part of smaller scales objects, trapped
in the larger one, disappear due to the tidal interactions [14]. This
process lead to effective homogenization of density distribution
inside the core $r\leq r_c$ (3). The third factor is the effect of
barionic body on neutralino distribution in its surrounding. Taking
these factors into account, one can approximately present the density
distribution for a small scale objects in a form:
\begin{equation}
\rho=\left\{
\begin{array}{cc}
\rho_0 &\qquad \qquad 0<r<r_c \\
\rho_0 \left(\frac{r}{r_c}\right)^{-\alpha} &\qquad \qquad r_c<r<R_x \\
 0 &\qquad \qquad  r>R_x \\
\end{array}
\right.
\end{equation}
We neglect here the changes of the density in the vicinity of barionic
body, as it  significantly affects only a small part ($\leq10^{-6}$)
of neutralino.

Density $\rho_0$ is connected with the full mass of the object $M_x$
by the relation:
\begin{equation}
M_x\approx\frac{4\pi}{3-\alpha}\rho_0\,R_x^3 \left[
\left(\frac{r_c}{R_x}\right)^{\alpha}-\frac{\alpha}{3}\left(\frac{r_c}{R_x}
\right)^3\right]
\end{equation}
From (5) it follows, that
\begin{equation}
\rho_0\approx\frac{3-\alpha}{4\pi}\,M_x\,R_x^{\alpha-3}\,r_c^{\alpha}
\end{equation}
and approximate density distribution is
\begin{equation}
\rho=\left\{
\begin{array}{cc}
\frac{3-\alpha}{4\pi}\,\frac{M_x}{R_x^3}\left(\frac{r}{r_c}\right)^{-\alpha} &
\qquad \qquad r_c<r<R_x \\
 0 &\qquad \qquad  r>R_x,\qquad r<r_c \\
\end{array}
\right.
\end{equation}
This model will be used below in the calculations of gamma radiation.

\section{Gamma radiation by neutralino stars}.

{\bf Extragalactic diffuse gamma radiation.}

Two main sources determine the diffuse gamma ray flux
$I_{\gamma}$.
First is an extragalactic radiation, which comes in integrated
form from all other galaxies of the Universe
$I_{1\gamma}$.
The second one is the gamma flux generated in our Galaxy
$I_{2\gamma}$:
\begin{equation}
I_{\gamma}=I_{1\gamma}+
I_{2\gamma}.
\end{equation}
We will determine below both these sources supposing that
$\gamma$-rays are generated by NeS.

Extragalactic radiation is determined directly by the full
energy losses in the Universe due to annihilation of neutralino
in NeS:
\begin{equation}
I_{1\gamma}=\frac{c}{4\pi\sqrt{3}}t_0 \dot{\varepsilon}_h \alpha_{\gamma}
\quad (cm^{-2} s^{-1} sr^{-1})
\end{equation}
Here $I_{1\gamma}$ is a number of $\gamma$-photons per 1 $cm^2$
per second, per steradian, $c/\sqrt{3}$- angle averaged velocity of photons,
$\dot{\varepsilon}_h$ - average energy losses per second in 1 $cm^3$
released in annihilation, $\alpha_{\gamma}$ - transformation
coefficient of dissipated energy into the number of gamma
quantum, $t_0$ is the lifetime of the Universe up to
redshift $z\approx 1$.

Energy losses
\begin{equation}
\dot{\varepsilon}_h = c^2 \rho_c \Omega f \tau^{-1}
\end{equation}
Here $\rho_c$ is the critical density of the Universe determined by
the Hubble constant
\begin{equation}
\rho_c=\frac{3H^2}{8\pi G}
\end{equation}
$\Omega$ is the dynamic mass parameter, $f$ - the parameter
characterizing the fraction of dark matter in the Universe
trapped in NeS, $\tau$ is the lifetime of neutralino in NeS:
\begin{equation}
\tau=\frac{N_x}{|\dot{N_x}|}
\end{equation}
Here $N_x$ - full number of neutralino in NeS
$$
N_x=\frac{M_x}{m_x}
$$
$M_x$ is mass of NeS, $m_x$  - mass of neutralino and $\dot{N_x}$ -
the neutralino annihilation rate:
\begin{equation}
\dot{N_x}= - 4\pi<\sigma v> \int n_x(r)^2 r^2 dr
\end{equation}
Here $n_x(r)$ is the number density of neutralino in NeS (7)
\begin{equation}
n_x(r)=\rho/m_x
\end{equation}
and $<\sigma v>$
is averaged multiplication of annihilation
crossection $\sigma$ and particle  velocity $v$. From (13),(14),(7)
it follows that
\begin{equation}
|\dot{N_x}|=\frac{(3-\alpha)^2}{4\pi(2\alpha-3)}\,\left(\frac{M_x}{m_x}
\right)^2 \, \frac{<\sigma v>}{R_x^3}\,
\left[\left(\frac{R_x}{r_c}\right)^{2\alpha-3}-1\right],
\end{equation}
and
\begin{equation}
\tau=\frac{4\pi(2\alpha-3)}{(3-\alpha)^2}\,\frac{m_x}{M_x}
 \, \frac{R_x^3}{<\sigma v>}\,
\left[\left(\frac{R_x}{r_c}\right)^{2\alpha-3}-1\right]^{-1}
\end{equation}
So, the flux of extragalactic gamma radiation $I_{1\gamma}$ is:
\begin{equation}
I_{1\gamma}=\frac{\sqrt{3}}{32\pi^2}\, \frac{\alpha_{\gamma}f\,\Omega
c^3H^2t_0}{G\tau}
\end{equation}

{\bf Galactic diffuse gamma radiation.}

Gamma radiation generated by NeS in our Galaxy depend on
direction:
\begin{equation}
I_{2\gamma}(\theta, \phi)=\frac{1}{4\pi}
\dot{N_{\gamma}}
\int_{0}^{R_h} n({\bf r})\, dr
\end{equation}
where $\dot{N_{\gamma}}$ is the number of gamma photons produced
per 1 sec.  in one NeS and $n({\bf r})$ -- the number density
of NeS in our Galaxy. According to [10]
\begin{equation}
n({\bf r})=\frac{3-\alpha}{4\pi} \frac{N_s}{R_h^3}
\times\left(\frac{|{\bf r}-{\bf r_s}|}{R_h}\right)^{-\alpha}
\qquad \qquad \alpha\approx 1.8
\end{equation}
where ${\bf r_s}$ is the coordinate of the Sun, $N_s$ is
the full number of NeS,
\begin{equation}
N_s=f_h\frac{M_{dh}}{M_x}
\end{equation}
and $M_{dh}$ is a full mass of dark matter in the halo, $R_h$ --
the size of halo, $f_h$ - parameter characterizing the fraction
of the halo dark matter trapped in NeS.

Choosing the axis of coordinate system along the direction
of vector ${\bf r_s}$ one can see from (18),(19), that the flux
$I_{2\gamma}$ does not depend on angle $\phi$. The dependence
on angle $\theta=arccos\,({\bf rr_s}/rr_s)$ is described by the integral
\begin{equation}
J=\int_{0}^{R_h}R_{h}^{\alpha}\left(r^2+r_s^2-2rr_s cos\theta
\right)^{-\alpha} dr \approx\frac{R_h^{\alpha}}{r_s^{\alpha-1}}
F(\theta)
\end{equation}
\begin{equation}
F(\theta)=\int_{0}^{\infty}\left(x^2-2x  cos\theta+1\right)^{-\alpha} dx
\end{equation}
In the last integral we took into account, that $R_h/r_s \gg 1$.
 A simple approximate expression for integral (22) could be obtained if we
take $\alpha=2$
\begin{equation}
F(\theta)= \frac{\pi-\theta}{sin\,\theta}
\end{equation}
One can show that approximation (23) coincide with (22) within a few percent
accuracy.

Taking into account that
\begin{equation}
\dot{N_{\gamma}}=\alpha_{\gamma} m_x c^2 |\dot{N_x}|
\end{equation}
we obtain from (13)-(22):
\begin{equation}
I_{2\gamma}=
I_{2\gamma}^0 F(\theta)
\end{equation}
were $I_{2\gamma}^0$ is the flux $I_{2\gamma}$ in anticenter direction
$\theta=\pi$
\begin{equation}
I_{2\gamma}^0=\frac{3-\alpha}{16\pi^2}\alpha_{\gamma}f_h
\frac{M_{dh}c^2}{R_h^2 \tau} \left(\frac{R_h}{r_s}\right)^{\alpha-1}.
\end{equation}
It is of considerable interest the relationship $p_{\gamma}$ between
Galactic $I_{2\gamma}$ and extragalactic $I_{1\gamma}$ diffusion fluxes.
From (17),(26) it follows:
\begin{equation}
p_{\gamma}=\frac{I_{2\gamma}^0}{I_{1\gamma}}=
\frac{\sqrt{3}(3-\alpha)}{4\pi}
\,\left(\frac{f_h}{f}\right)
\,\frac{M_{dh}}{\Omega\rho_cR_h^2ct_0}
\left(\frac{R_h}{r_s}\right)^{\alpha-1}
\end{equation}

{\bf Discrete sources of gamma radiation}.

NeS could be considered as discrete source of gamma
radiation also. Let us determine the intensity of these sources.
Formulae (24) gives the full number of
gamma photons produced per 1 sec. in one NeS.
It follows from (24) that the intensity of gamma radiation from
one NeS at the distance $r_x$ from the source is
\begin{equation}
I_{\gamma}=\frac{1}{4\pi}\alpha_{\gamma}m_xc^2|\dot{N_x}|r_x^{-2}
\end{equation}
As the NeS density in the vicinity of the Sun according to (4) is
\begin{equation}
n_s=\frac{3-\alpha}{4\pi}\,\frac{N_s}{R_h^3}\left(\frac{R_h}{r_s}
\right)^{\alpha}
\end{equation}
it follows that with the probability of the order of unity one can
see several NeS as a source of high energy gamma quants
 with intensities
\begin{equation}
I_{\gamma c}=\frac{(1-\alpha/3)^{2/3}}{4\pi}\alpha_{\gamma}
f_h^{2/3}c^2\frac{M_{dh}^{2/3} M_x^{1/3}}{\tau R_h^2}
\left(\frac{R_h}{r_s}\right)^{2\alpha/3}
\end{equation}
So $I_{\gamma c}$ gives the characteristic minimal sensitivities
for the devices capable to observe discrete NeS by its gamma radiation.

{\bf Galaxies as a spread sources of gamma radiation.}

The intensity of gamma radiation generated by NeS in the galaxy M31
(Andromedae) should be described by equations (18)-(19). The only difference
that $r_s$ should be changed for $r_{sa}$- the coordinate of Sun in the
reference system connected with the centrum of Andromedae. In
integration $dr$ it is necessary to bear in mind
that the density of NeS in Andromedae is:
\begin{equation}
n_a({\bf r})=\frac{3-\alpha}{4\pi}\,\frac{N_s}{R_{ha}^3}
\left(\frac{|{\bf r}-{\bf r}_{sa}|}{R_{ha}}\right)^{-\alpha}
\Theta\left(1-\frac{|{\bf r}-{\bf r}_{sa}|}{R_{ha}}\right)
\end{equation}
where $\Theta$ is a theta function which changes the limits in
the integral (18). Taking into
account that dimension of Andromedae halo $R_{ha}$ is less than $r_{sa}$
we obtain from (18),(31):
\begin{equation}
I_{\gamma}^A=\frac{2(3-\alpha)^3}{(4\pi)^3(2\alpha-3)}\,
\frac{\alpha_{\gamma}f\, c^2M_{dh}^AM_x r_{sa}^{1-\alpha}}
{m_xR_{ha}^{3-\alpha}}\,
\frac{<\sigma v>}{R_x^3}
\left[\left(\frac{R_x}{r_c}\right)^{2\alpha-3}-1\right] K(\theta)
\end{equation}
$$
K(\theta)\approx\frac{1}{|sin\,\theta|^{\alpha-1}}
arctg\sqrt{\frac{R_{ha}^2}{r_{sa}^2 sin^2\theta}-1}
$$
Here $\theta$ is the angle between any given direction and direction
from the Sun to the centrum of M31. We see that the gamma radiation
from M31 is spread over the range of the angles $\theta\approx
R_{ha}/r_{sa}\sim 20^o$. To obtain the full intensity of gamma radiation
of Andromedae $I^A$ we have to integrate over $\phi$ and $\theta$ angles.
Taking into account that $(R_h/r_{sa})^2\sim 0.3 < 1$, we obtain
\begin{equation}
I^A\approx\frac{1}{16\pi^2}\alpha_{\gamma}f\frac{M_{dh}^A M_x c^2}
{m_x r_{sa}^2}\, \frac{<\sigma v>}{R_x^3}\left[
\left(\frac{R_x}{r_c}\right)^{2\alpha-3}-1\right]
\end{equation}

In the case of LMC it is naturally
to suppose that LMC has lost most part of its
dark matter halo due to the
interaction with our Galaxy. Because of this reason  the full mass
of dark matter in LMC could be of the same order or even less than
the barionic mass, so
$$
M_d^{LMC}\leq M_b^{LMC}
$$
In such conditions distribution (19),(31) does not relates to the
dark matter of LMC. So the LMC should be seen as a spread source of
gamma radiation with the full intensity (33), only $r_{sa}$, $M_{dh}^A$,
$R_{ha}$ should be changed to the analogous characteristics of the LMC.

Gamma radiation from other nearest galaxies is described by the same
equations as for Andromedae
(32),(33). The distant galaxies should be seen as point-like and their
radiation is given by formulae (33).

\section{Comparison with observations.}

{\bf Diffuse gamma radiation}.

To compare with observational data one has to determine the main
parameters. We will take $H\approx 70 km/s\,Mpc$ and for NeS we will use:
\begin{equation}
 M_x=0.5M_{\odot},\quad R_x\approx 4\times 10^{14} cm \quad
m_x=10\, Gev
\end{equation}
Neutralino annihilation process could be described by simple
expression [4]:
\begin{equation}
<\sigma v>=
<\sigma v>_0 \times \frac{r_g}{R_x}\times
\left(\frac{m_x}{10 \, Gev}\right)^2 \qquad \qquad
<\sigma v>_0\approx \left(10^{-26} - 10^{-27}\right)
cm^3 s^{-1}
\end{equation}
where $r_g=2GM_x c^{-2}$ is the gravitational radius of NeS.
Then the neutralino lifetime $\tau$ (16) is:
\begin{equation}
\tau=5.4\times 10^{24}\left(\frac{10\,Gev}{m_x}\right)
\left(\frac{R_x}{4\times 10^{14}cm}\right)^4
\left(\frac{10^{33}\,g}{M_x}\right)^2
\left(\frac{10^{-27} cm^{3}s^{-1}}{<\sigma v>_0}\right) \,\, s
\end{equation}
The total number of photons generated by NeS (24) is
\begin{equation}
\dot{N_{\gamma}}=
1.03\times 10^{32}\alpha_{\gamma}
\left(\frac{M_x}{10^{33}g}\right)^3
\left(\frac{m_x}{10\,Gev}\right)
\left(\frac{4\times10^{14}cm}{R_x}\right)^4
\left(\frac{<\sigma v>_0}{10^{-27}}\right)\,\, \frac{photon}{ s}
\end{equation}
Here $\alpha_{\gamma}$ describes the number of photons
with energy $E_{\gamma}>100$Mev on 1 Gev of energy
dissipated in neutralino annihilation.
The diffuse extragalactic gamma radiation is:
\begin{equation}
I_{1\gamma}=1.4\times 10^{-4}\alpha_{\gamma}\Omega
\left(\frac{f}{0.5}\right)
\left(\frac{t_0}{2\times 10^{17}s}\right)
\left(\frac{m_x}{10\,Gev}\right)
\end{equation}
$$
\times\left(\frac{\rho_c}{10^{-29}\,g\,cm^{-3}}\right)
\left(\frac{4\times10^{14}cm}{R_x}\right)^4
\left(\frac{M_x}{10^{33}g}\right)^2
\left(\frac{<\sigma v>_0}{10^{-27}}\right)
\frac{photon}{cm^2 s\,str}
$$
Diffuse Galactic gamma radiation depends on the full mass of the
dark matter of galaxy $M_{dh}$ and the size of the halo $R_h$ (26).
We will take as usual [10]:
\begin{equation}
M_{dh}=2\times 10^{12} M_{\odot},\qquad R_h=200 kpc, \qquad
r_s=8.5 kpc
\end{equation}
Then the relation of components (27) is
\begin{equation}
p_{\gamma}=\frac{I_{2\gamma}^0}{I_{1\gamma}}= \frac{0.36}{\Omega}
\left(\frac{f_h}{f}\right)
\left(\frac{M_{dh}}{2\times10^{12}M_{\odot}}\right)
\left(\frac{200\,kpc}{R_h}\right)^{2-\alpha}
\left(\frac{10^{-29}}{\rho_c}\right)
\left(\frac{2\times10^{17}s}{t_0}\right)
\end{equation}
The minimal Galactic diffusive flux is
\begin{equation}
I_{2\gamma}^0=6.1\times 10^{-5}\alpha_{\gamma}
\left(\frac{f_h}{0.5}\right)
\left(\frac{M_{dh}}{2\times10^{12}M_{\odot}}\right)
\left(\frac{200\,kpc}{R_h}\right)^{2-\alpha}
\end{equation}
$$
\times\left(\frac{m_x}{10\,Gev}\right)
\left(\frac{4\times10^{14}cm}{R_x}\right)^4
\left(\frac{M_x}{10^{33}g}\right)^2
\left(\frac{<\sigma v>_0}{10^{-27}}\right)\, \frac{photon}{cm^2 s\,str}
$$
It is easy to see that
gamma background $I_{b\gamma}$ -- the minimal flux, which could be
observed near the Sun is
\begin{equation}
I_{b\gamma}=I_{1\gamma}+I_{2\gamma}^0=I_{1\gamma}(1+p_{\gamma})
\end{equation}
and the full diffuse flux of gamma radiation:
\begin{equation}
I_{\gamma}=I_{1\gamma}+I_{2\gamma}^0 F_{\theta}
=I_{1\gamma}\left(1+p_{\gamma}F({\theta})\right)
\end{equation}
where function $F(\theta)$ is given by equations (22),(23).

Let us compare now the theory with the observational data.
We will consider fluxes of gamma photons with energies $E>100$Mev.
Gamma background for these energies is [15,16,23]:
\begin{equation}
I_{b\gamma}\approx 1.5\times 10^{-5}\, \frac{photon}{cm^2 s\,str}
\end{equation}
This value could be obtained from (38),(42) if parameter
\begin{equation}
q=\alpha_{\gamma}
\left(\frac{f}{0.5}\right)\Omega
\left(\frac{m_x}{10\,Gev}\right)
\left(\frac{4\times10^{14}cm}{R_x}\right)^4
\left(\frac{M_x}{10^{33}g}\right)^2
\left(\frac{<\sigma v>_0}{10^{-27}}\right)
\left(\frac{t_0}{2\times 10^{17}s}\right)\approx 0.1
\end{equation}
For the same parameter $q$ and $p=0.25$. the longitudinal dependence of
the full flux $I_{\gamma}(\theta)$  (43),  is compared at Fig.1
with the observational data. An agreement between the theory
and observations is seen.

It should be emphasized, that formulae (35) for $<\sigma v>$  follows
from the supposition, that the dominant annihilation process of
neutralino proceed as p-wave [4]. Some authors (see [8])
considered for neutralino particles $s$-wave annihilation also.
The existence of $s$-wave lead in our case to annihilation crossection
$<\sigma v>$ much higher -- up to 5 orders of magnitude. We see,
that such particles give too strong gamma flux and so they could
not be considered as the dark matter particles for NeS. This
conclusion agrees with the recent statement
[17]\footnote{Note that in [17] was considered a very specific
model of NeS. The authors neglected the existence of barionic core
and supposed the density of neutralino in active region to be extremely
high: $n_x\approx5\times10^{33} cm^{-3}$ more than 20 orders of magnitude
higher than follows from (3),(7). It leads to the neutralino lifetime
$\tau_x\approx10^{-4} s$ and naturally results in
nonrealistic level of the flux of gamma radiation}.
It was also clearly formulated previously in [4]
 from the analysis of NeS lifetime.

So we see, that the comparison of the theory with the observations
of diffuse gamma radiation gives significant constrains on the type
of neutralino particles in NeS: it should be only particles with
strongly dominating $p$-wave annihilation process. For example
neutralino particles of such type -- light photino, were recently
considered in   [18],[19].
If dark matter consist of this type particles, the
existence of NeS is in  agreement with observations
of diffuse flux. In
the following section we will give other impressive
consequences of this statement.

{\bf Discrete sources of gamma radiation}.

Discrete sources of high energy gamma radiation $E\geq 100\,Mev$
are partially identified with active galactic nucleus, pulsars and
other active objects in the Galaxy. Nonidentified sources were discovered
 by COS B [20], mostly effectively they are studied by EGRET
installation at COMPTON observatory.

The recently published second EGRET catalog [21] has 71 sources
nonidentified with known objects. They are classified according
to the peculiarities of observational morphology into three groups:
$em$-- possibly extended or multiple sources, $C$-- source confusion
that may affect flux, significance, position,etc. and definite point-
like sources in the EGRET observational $1^o$ diagram ($P$).

As the sources of gamma radiation by NeS we can consider only the
last group. This group of 32 sources is listed in the Table 1.
Averaged flux of photons
$S$ and its dispersion $\sigma$ in units $10^{-8}photon/cm^2\,s$
are obtained from
EGRET catalog by direct averaging of presented data. The position of
$P$ sources in the sky in galactic coordinates $l$ and $b$ is shown
at the Fig.2. Note that we do not show at the figure five last $P$
sources from the Table since their average intensity $S\geq80$ seems
to be too high for NeS (see below). As is seen from the figure
the distribution of  the $P$ sources in the sky looks quite
isotropic. The small anisotropy (for example small excess of the
sources in the anticenter region $l\sim 180^o$) could be connected
with excess of the EGRET exposure time (see Fig.1 in [21]). In the same
time some deficit of the $P$ sources in the central region which is
seen at the figure could be the result of a strong growth of the
diffuse flux in this region (see Fig.1). Because of this mainly
intensive point-like sources could be seen in this direction. And
really, among the five omitted strongest sources two belongs
 to the central region  and two -- to
the Galaxy plane. Note, that in principal among NeS could exist
quite strong sources as they could be
 close enough to the observer. But in
our case their definite concentration to the central region and
Galaxy plane make them to be suspicious as a NeS sources.

So we see that the distribution of nonidentified $P$-sources at the
sky is fairly well isotropic. $log\,N-log\,S$ diagram for this sources
is shown at the Fig.3. It confirms well enough not only isotropy, but
a homogeneous distribution of the sources in the sky
around the observer also. All the deviations from -3/2 law are
well inside the statistical errors. Only first  points could be
considered as suspicious (they are signed by $\odot$ at the Fig.2.).

The isotropy and
$log\,N-log\,S$ curve
is in a full agreement with the supposition that a part
of nonidentified P-sources of gamma radiation consist of NeS.
$log\,N-log\,S$ curve determine the characteristic intensity of the
source at the distance
$$
r_0=\left(\frac{3}{4\pi\, n_s}\right)^{1/3}
$$
where $n_s$ is the density of NeS in the vicinity of the Sun.
In our case this intensity according to Table 1 and Fig.2 is
\begin{equation}
I_0\approx5\times 10^{-7} \, \frac{photon}{cm^2 s}
\end{equation}
Let us emphasize now that according to (30),(37) the same intensity is given
by formulae (30):
\begin{equation}
I_0=\frac{\dot{N_{\gamma}}}{4\pi r_0^2} = 10^{-7}\alpha_{\gamma}
\left(\frac{M_x}{10^{33}g}\right)^{7/3}
\left(\frac{m_x}{10\,Gev}\right)
\left(\frac{4\times 10^{14}cm}{R_x}\right)^4
\left(\frac{<\sigma v>_0}{10^{-27}}\right)
\end{equation}
$$
\times\left(\frac{M_{dh}}{2\times 10^{12} M_{\odot}}\right)^{2/3}
\left(\frac{200\,kpc}{R_h}\right)^{0.8}
\left(\frac{f}{0.5}\right)^{2/3}
\frac{photon}{sm^2 s}
$$
Taking the value for $\alpha_{\gamma}m_x<\sigma v>_0$
 from (45) we determine the average source flux value $I_0$
\begin{equation}
I_0= 10^{-8}\frac{1}{\Omega}
\left(\frac{f}{0.5}\right)^{-1/3}
\left(\frac{2\times 10^{17}s}{t_0}\right)
\end{equation}
$$
\times\left(\frac{M_x}{10^{33}g}\right)^{1/3}
\left(\frac{M_{dh}}{2\times 10^{12}M_{\odot}}\right)^{2/3}
\left(\frac{200\, kpc}{R_h}\right)^{0.8}
\frac{photon}{sm^2 s}
$$
Note, that this value is in an agreement with the diffuse flux (42),
(43), if we consider this flux gathered from discrete sources. On the other
hand one can see a discrepancy ($\sim 50$) comparing (48) with the
intensity of directly observed $P$-sources (46).
Several factors could be pointed out which explain or diminish this
discrepancy. First, not all P-sources are supposed to be NeS and so
it is natural to expect, that the strongest part of the sources
(in Fig.3) is not NeS.
Second, we did not take into account the existing
 dispersion of NeS on masses $M_x$ and dimensions $R_x$. Because of this
 it could be, that a part of the observed sources belongs not to
the main bulk but to the mostly intensive fluctuations (what means
those, having the highest $M_x$ and the lowest $R_x$). Third, a small
change of parameters: diminishing
of $\Omega, R_h, t_z$ or growth of $M_{dh}, R_x$ and $M_x$
can diminish the discrepancy also. So, being significant the discrepancy
between (46) and (48) does not seem as a decisive one and we can speak
 about relative agreement between
diffuse and discrete approaches in the description of gamma radiation
produced by neutralino annihilation in NeS.

{\bf Galaxy M31 as a spread source of gamma radiation}.

To determine the intensity of Andromedae gamma radiation we will use
the following values of main parameters
$$
M_{dh}^A=3\times 10^{12} M_{\odot}, \qquad R_{ha}=250\,kpc
\qquad r_{sa}=600\, kpc
$$
Then as follows from (32) the flux from Andromedae is:
$$
I_{\gamma}^A\approx
1.1\times 10^{-5} K({\theta}) \alpha_{\gamma}
\left(\frac{f}{0.5}\right)
\left(\frac{M_{dh}^A}{3\times 10^{12}M_{\odot}}\right)
$$
\begin{equation}
\times\left(\frac{M_x}{10^{33}g}\right)^2
\left(\frac{4\times10^{14}cm}{R_x}\right)^4
\left(\frac{250\,kpc}{R_{ha}}\right)^2
\left(\frac{m_x}{10\,Gev}\right)
\left(\frac{<\sigma v>_0}{10^{-27}}\right)\,
\frac{photon}{cm^2 s\,str}
\end{equation}
$$
K({\theta})=\frac{1}{|sin\,\theta|^{\alpha-1}} \,
arctg \sqrt{\left(\frac{0.416}{sin\,\theta}\right)^2-1}
$$
It means, that for the EGRET receiving system having the width
$1^o$ the flux $I_{\gamma}$ from the centrum of Andromedae would be
$$
I_{\gamma}^A=1.3\times 10^{-7} \alpha_{\gamma}<\sigma v>_0
\left(\frac{m_x}{10\,Gev}\right)\, \frac{photon}{cm^2 s}
$$
Using now the relation (45) we obtain, finally the flux in EGRET
from the centrum:
\begin{equation}
I_{0\gamma}^A\approx1.3\times 10^{-8}\, \frac{photon}{cm^2 s}
\end{equation}
That is the flux from the galaxy Andromedae itself (it has
approximately $1^o$ dimension).
This flux is slightly lower than the usual EGRET sensitivity.
Moving out from the galaxy centrum to the halo ($\theta>1^o$)
the flux from Andromedae halo , as follows from (32) is diminishing
\begin{equation}
I_{\gamma}^A(\theta)=
I_{0\gamma}^A\times \left(\frac{1^o}{\theta^o}\right)^{0.8}
\end{equation}
and goes to zero at $\theta\approx 20^o$. The  integral flux from
Andromedae halo is (33)
\begin{equation}
I_{\gamma}\approx2.0\times 10^{-6}
\alpha_{\gamma}
\left(\frac{f}{0.5}\right)
\left(\frac{M_{dh}^A}{3\times 10^{12}M_{\odot}}\right)
\left(\frac{M_x}{10^{33}g}\right)^2
\end{equation}
$$
\times\left(\frac{4\times10^{14}cm}{R_x}\right)^4
\left(\frac{m_x}{10\,Gev}\right)
\left(\frac{<\sigma v>_0}{10^{-27}}\right)\,
\frac{photon}{cm^2 s}
$$
Using again relation (45) we find the total flux from Andromedae halo
$$
I_{\gamma}^A\approx2\times 10^{-7}\,\frac{photon}{cm^2 s}
$$
what means, that in principle it could be observed by EGRET.

\section{Conclusion}.

Let us summarize the main results of the paper.

1. A new mechanism for generation of gamma radiation (GR) connected
with neutralino annihilation in noncompact dark matter objects
-- Neutralino Stars (NeS) is proposed.

2. On the basis of microlensing data and the theory of NeS
the diffuse flux both of extragalactic and Galactic origin is
calculated. The theory is shown to be in a reasonable agreement with
observations. We especially note that this statement does not
mean, that other sources of gamma radiation (for example cosmic
rays) are not essential: first of all, NeS do not affect
nonsphericaly symmetrical component of Galactic GR [22] and second,
 it is quite possible to make small changes
in the parameters, which will give a free space for conventional
mechanisms of gamma production.

3. From the EGRET catalog of nonidentified sources the point-like
sources of gamma radiation (P) are separated. They are proved
to be distributed isotropically and homogeneously in space
(-3/2 law for $log\,N-log\,S$ curve).

We suppose that a part of P- sources observed recently by EGRET
is the gamma radiation of individual NeS. So, the new interpretation
of nonidentified gamma sources is proposed. An agreement
between the theory and observations is demonstrated.

It is shown that diffuse gamma radiation being considered as
partialy produced
by NeS  is in a relative agreement with EGRET observations of P-
sources.

4. Of course, more detailed studies of gamma radiation and comparison
with the predictions of the theory are needed to prove that NeS are
really observed and that NeS really
affect significant features of the existing
gamma radiation. Let us mention some of them:

a) The theory predicts that  new discovered nonidentified
point-like P- sources especially in the low intensity range
  should be isotropically distributed
and should obey -3/2 law for $log\,N-log\,S$ curve.

b) The theory predicts the existence of a spread gamma radiation
from Andromedae with quite definite features. The discovery  and
detailed study of this source would mean the direct observation
of giant dark matter halo of the galaxy and would establish that
this halo consist of NeS.

c) The studies of gamma spectrum of the P-sources is of a great interest.
The theory predicts that this spectrum should be the same for all
sources and that it is fully determined by the process of neutralino
annihilation, which lead mostly to the $\pi^o$ decay gamma spectrum.

d) Since some of observed P-sources should be on the distances of the
order $10^{18}$cm from the Sun, it may be possible to observe in
gamma rays their regular motion, which should be of the order $3^{'}$
per year and one year oscillations with amplitude $\sim3^{''}$.

e) According to the theory P- sources and MACHO objects are the same.
It has a special interest to combine gamma and microlensing observations
of the same objects. Microlensing is not usual one: as the distances
to the object is only (1-10)pc, the Einstein radius is of the order
$(1-3)10^{12}$cm and characteristic timescale of microlensing event
is (0.3-1)day.

5. If the hypothesis that NeS are an effective sources of GR would
be confirmed, then
the presented comparison of the theory with observations gives
fundamental information about CDM particles:

a) Particles should annihilate like Majorana particles.

b) A dominant annihilation process is going in $p$-wave chanel.
$s$-wave chanel is practically completely suppressed.

c) A special relation (45) characterizing parameters of the particles
is established. It connects the mass of the particle ($m_x$), annihilation
crossection ($<\sigma v>$) and coefficient of gamma quants production
due to annihilation ($\alpha_{\gamma}$).

The further theoretical and experimental studies of these
problems are of fundamental interest.

\vspace{0.5cm}

The authors are grateful to V.L.Ginzburg for permanent interest and
fruitful discussion, and to
I.Axford, V.A.Dogiel,
A.Lykyanov and V.Sirota for valuable remarks.

This research was supported by  the Russian Fund of Fundamental Research \\
(Grant  96-02-18217).

\vspace{1cm}

{\bf References}

\vspace{1cm}

1. C.Alcock et all Nature 365 (1993) 621.

2. M.R.Pratt et al (The MACHO collaboration) preprint astro-ph/9606134.

3. A.V.Gurevich and K.P.Zybin, Phys.Lett.A 208 (1995) 267.

4. A.V.Gurevich,K.P.Zybin,V.A.Sirota, Phys.Lett.A 214 (1996) 232

5. A.V.Gurevich,K.P.Zybin,V.A.Sirota, II Int.Sakharov Conference, Moscow
(1996)

6. A.V.Gurevich,K.P.Zybin,V.A.Sirota, VIII Conference on neutrino, dark matter

and Universe Blois, (1996)

7. L.Roskowski, Phys.Lett.B 278 (1992) 147.

8. A.Botino et al Astroparticle Phys. 2 (1994) 77

9. J.P.Ostraiker  Ann.Rev.of Astron.Astroph. (1992) 689.

10. A.V.Gurevich and K.P.Zybin Physics Uspekhi 38 (1995) 687

11. A.V.Gurevich and K.P.Zybin Sov.Phys.JETP 67 (1988) 1957

12. A.V.Gurevich,K.P.Zybin,V.A.Sirota, Phys.Lett.A 207 (1995) 333.

13. A.V.Gurevich, M.I.Zelnikov and K.P.Zybin Sov.Phys.JETP 107 (1995) 321.

14. A.V.Gurevich,K.P.Zybin,V.A.Sirota, Physics Uspekhi (in press).

15. Astrophysics of Cosmic Rays (ed. V.L.Ginzburg) Elsevier Science
Publisher B.V.,1990.

16. C.E.Fichtel,G.A.Simpson and D.J.Thompson Astrophys.J. 222 (1978) 833.

17. V.Berezinsky, A.Botino, G.Mignola preprint CERN-TH/96-283 October 1996.

18. G.R.Farrar and E.W.Kolb Phys.Rev.D 53 (1996) 2990

19. E.W.Kolb and A.Riotto, astro-ph/9601096

20. G.F.Bignami and W.Hermsen, Ann.Rev.Astron.Astrophys. 21 (1983) 67.

21. D.J. Thompson et al Astrophy.J.Suppl. 101 (1995) 259.

22. S.D.Hunter et al Astron. and Astrophys. (accepted for publication).

23. D.A.Kniffen et al Astron. and Astrophys. (accepted for publication).

\newpage
{\bf Figure caption.}

\vspace{1cm}

Fig.1. Intensity distribution of gamma radiation in the direction

of galactic centre as a function of galactic latitude. The crosses

and hystogram denote the obsered values [15], curve -- the theory (43).

\vspace{0.5cm}

Fig.2. Sky distribution of P-sources.

\vspace{0.5cm}

Fig.3. $log\,N-log\,S$ distribution of P-sources compared with
-3/2 law.

\vspace{0.5cm}

Table.1. Point-like (P) sources. Averaged intensity $S$
and dispersion  $\sigma$ in units $10^{-8}\frac{photon}{cm^2\,s}$

\begin{table}[p]
\caption{}
\begin{center}
\begin{tabular}{|c|r@{.}l|r@{.}l|}\hline
Name&\multicolumn{2}{|c|}{$S$}&\multicolumn{2}{|c|}{$\sigma$}\\ \hline
\makebox[5cm][c]{2EG J1054+5736}&\makebox[1cm][r]{7}&
\makebox[1cm][l]{6}&\makebox[1cm][r]{1}&\makebox[1cm][l]{2}\\
2EG J0403$+$3357&12&1&0&1\\
2EG J1233$-$1407&13&6&3&2\\
2EG J0159$-$3557&13&7&1&3\\
2EG J1239$+$0441&13&8&1&5\\
2EG J0119$+$0312&15&0&2&0\\
2EG J0216$+$1107&15&2&0&6\\
2EG J1136$-$0414&15&8&\multicolumn{2}{|c|}{---}\\
2EG J1346$+$2942&16&9&2&3\\
2EG J1731$+$6007&17&6&4&1\\
2EG J1332$+$8821&18&0&4&5\\
2EG J1457$-$1916&21&8&5&6\\
2EG J0323$+$5126&23&6&0&5\\
2EG J0545$+$3943&24&6&3&4\\
2EG J1248$-$8308&24&8&3&5\\
2EG J0744$+$5438&26&2&5&3\\
2EG J1430$+$5356&26&3&\multicolumn{2}{|c|}{---}\\
2EG J0852$-$1237&26&9&8&2\\
2EG J1950$-$3503&29&8&11&6\\
2EG J2006$-$2253&30&6&6&2\\
2EG J2354$+$3811&36&4&0&6\\
2EG J1821$-$7915&36&7&4&8\\
2EG J2227$+$6122&50&4&2&4\\
2EG J0008$+$7307&56&0&8&0\\
2EG J0618$+$2234&61&0&5&5\\
2EG J1835$+$5919&66&0&5&8\\
2EG J1049$-$5847&69&0&7&2\\
2EG J1443$-$6040&78&0&4&5\\
2EG J1528$-$2352&92&0&\multicolumn{2}{|c|}{---}\\
2EG J1021$-$5835&98&0&7&6\\
2EG J0241$+$6119&114&0&3&8\\
2EG J1746$-$2852&138&0&12&0\\ \hline
\end{tabular}
\end{center}
\end{table}

\end{document}